\documentclass[12pt]{article}
\usepackage{epsf}
\usepackage{graphicx}
\begin{document}
\begin{center}
{\bf\large{The preliminary result from spectra of $K^0_s \pi^-$ in
reaction p+propane at 10 GeV/c.}}

\vskip 2mm P.Zh.Aslanyan $^{1,2,\dag} $

 \vskip 3mm
{\small (1) {\it Joint Institute for Nuclear Research }
\\
(2) {\it Yerevan State University }
\\
$\dag$ {\it E-mail: paslanian@jinr.ru }}

  \end{center}

% ----------------------------------------------------------------

\begin{abstract}
 The experimental data from 2m propane bubble chamber have been analyzed
to search for  scalar meson $\kappa(800)$ in a $K^0_s\pi$ decay mode
for the reaction p+$C_3H_8$ at 10 GeV/c. The $K^0_s\pi^-$ invariant
mass spectrum has shown resonant structures with
$M_{K^0_s\pi^-}$=730, 900 and $\Gamma$=143, 48  MeV/$c^2$,
respectively. The statistical significance are estimated to be of
14.2$\sigma$ and 4.2$\sigma$, respectively. The peak in M(900) is
identified as reflection from the well known resonance with mass of
892 MeV/c$^2$.

\end{abstract}

% ----------------------------------------------------------------omtrica
\section{Introduction}

Already back in 1977 Jaffe~\cite{jaffe1}using the bag
model\cite{jaffe2} in which confined colored quarks and gluons
interact as in perturbative QCD, suggested the existence of a light
nonet composed of four quarks mesons\cite{kappa}. There are
theoretical arguments in favour of a light and broad
$\kappa(800)$\cite{kappa} pole near the $K\pi$ threshold. However,
the experimental evidence is not conclusive. The scalar mesons have
vacuum quantum numbers and are crucial for a full understanding of
the symmetry breaking mechanisms in QCD, and presumably also for
confinement. In a recent topical review \cite{close}, it was
suggested that the lightest scalars are at the central core composed
of a four quarks. Following Jaffe's QCD arguments this central core
would consist predominantly of a four-quark $(q,q){\overline{3}}(
\overline{q},\overline{q})_3$ state. At larger distances from the
core the four quarks would then recombine to a pair of colour
singlet $q\overline{q}$'s, building two pseudoscalar mesons as a
meson cloud at the periphery. Suggestions that the $\sigma(600)$ and
$\kappa(800)$ could be glueballs have been made. There are
theoretical arguments for why a light and broad
$\kappa(800)$($q\overline{q}$ or 4-quark state) pole can exist near
the $\kappa$ threshold and many phenomenological papers(Naive quark
model\cite{kappa}) support its existence. The $\sigma(600)$ and
$\kappa(800)$ indeed belong to the same family as the $f_0(980)$ and
$a_0(980)$ mesons (say if the $\sigma(600)$ were composed of 2 or 4
u and d type quarks) then no such mechanism would suppress the decay
$\sigma(600)\to \pi^+\pi^-$ or $\kappa(800)\to K\pi$.

The  PDG\cite{pdg}   numerous determinations of the pole mass in the
   neighbourhood of 600 MeV. There is believe that experimental
   groups should look for pole positions in their data analysis,
   which also include the aforementioned nonlinear effects from
   S-wave thresholds. The E791\cite{e791} Collaboration reported a light  ?
   with mass 797 MeV and width 410 MeV, but uses a Breit–Wigner amplitude.
    This claim was, however, not confirmed by the CLEO Collaboration.
     A lighter and very broad  $\kappa$ pole is nonetheless possible and should
     be looked for in future data analyzes\cite{kappa}.

\begin{figure}[ht]
 \centerline{\epsfysize=100mm
 \epsfbox{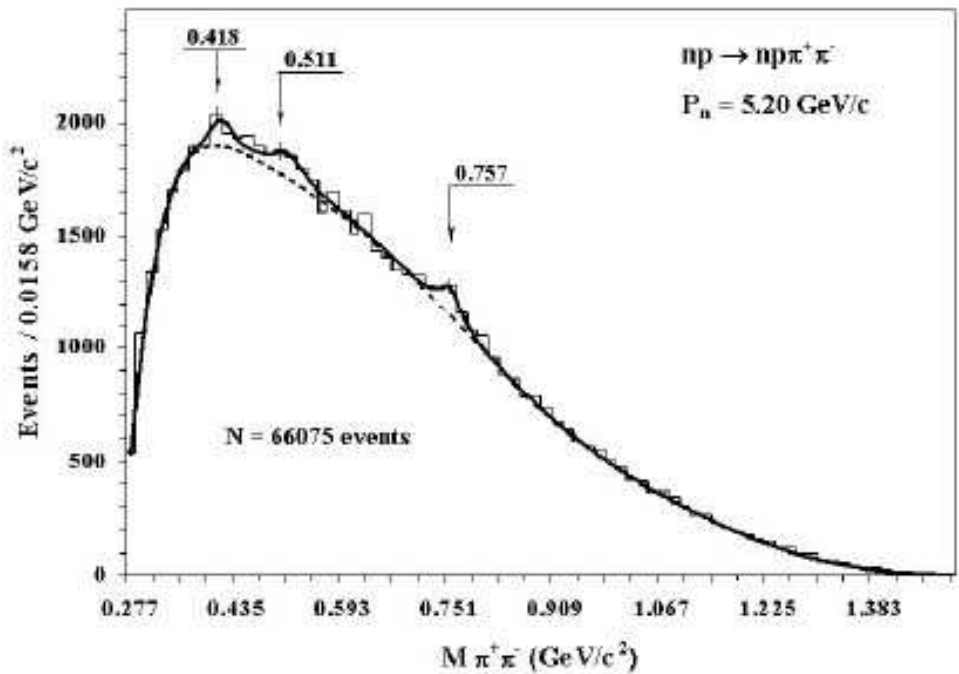}}
 \caption{The effective mass distribution of ($\pi^+,\pi^-$) -combinations
 from the total statistics of the reaction
np$\to \pi^-\pi^+$ np at  $P_n$ =5.2 GeV/c. The dotted curve is the
background taken in the form of a superposition of Legendre
polynomials up to the 10-th degree, inclusive.  } \label{troy1}
    \end{figure}

The effective mass distribution of $\pi^+\pi^-$ combinations from
total statistics of the reaction np$\to\pi^+\pi^-$X at $P_n$=5.2
GeV/c. The distribution is approximated by a polynomial background
curve and by 3 resonance curves taken in the Breit-Wigner form
Fig.~\ref{troy1}. At least 3 states with quantum numbers of
$\sigma_0$ –meson $0^+(0^{++})$ have been found at masses of 418,
511 and 757 MeV/$c^2$. The fact low -mass $\sigma_0$-mesons are
glueballs is one of the possible interpretations.

\section{Experiment}

The experimental information of more than 700000 stereo photographs
from the LHE JINR 2m bubble chamber \cite{penev}-\cite{aslan} were
used to select the events with $V^0$ strange particles.

\begin{figure}
%\centerline{
% \includegraphics[width=50mm,height=45mm]{fig1.ps}}
  \includegraphics[height=0.6\textheight]{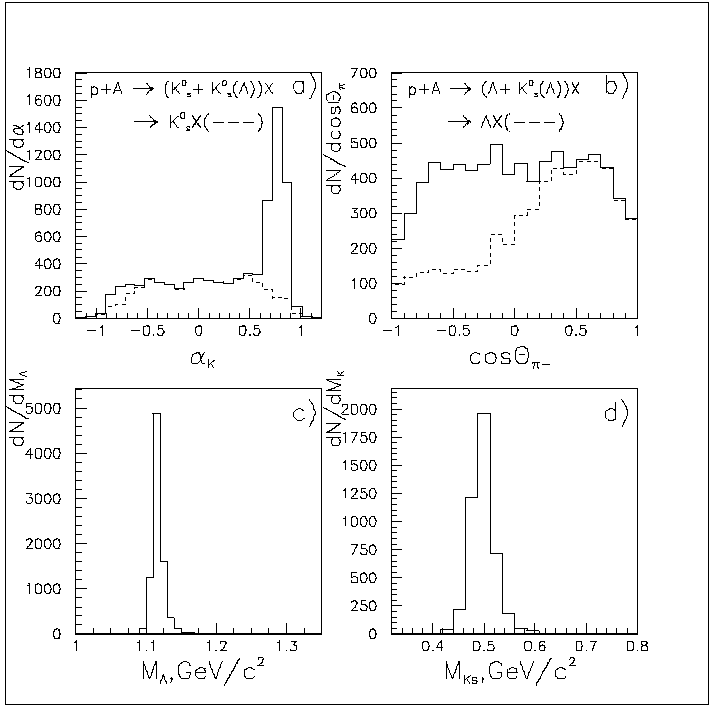}
  \caption{(a) and (b) distributions  of  $\alpha$ (Armenteros parameter) and  cos$\Theta^*$-
  are used  for  correctly identification  of   the undivided
  V0s. $\alpha = (P^+_{\parallel}-P^-_{\parallel})/((P^+_{\parallel}+P^-_{\parallel})$. Where $P^+_{\parallel}$
  and  $P^-_{\parallel}$ are the parallel components of momenta  positive and  negative
  charged  tracks. cos$\Theta^*$ - is the angular distribution  of $\pi^-$ from
  $K_s^0$  decay in rest frame $K_s^0$. Distributions of $\alpha$  and  cos$\theta$- were isotropic in the rest  frame
  of $K_s^0$  when  undivided $\Lambda K_s^0$ were assumed to be  events  as $\Lambda$.
  c) and (d) distributions of experimental $V^0$ events produced
 from interactions of beam protons with propane: c) for the effective mass of
  $M_{\Lambda}$; d) for the effective mass of $M_{K^0_s}$.\label{inv0}
 }
\end{figure}

The effective mass distribution of 8657-events of
$\Lambda$-hyperons, 4122-events of $K^0_s$-mesons were consistent
with their mass values from PDG\cite{pdg} (Figure~\ref{inv0}).   The
average geometrical weights were 1.34$\pm$0.03 for $\Lambda$ and
1.22$\pm$0.04 for $K^0$.

As one of backgrounds for experimental data analysis   was used
FRITTIOF model \cite{fri1} in collision p+C$\to K^0_s$X. The
experimental dates are described by the FRITIOF model satisfactorily
\cite{aslan}.

\section{$K^0_s \pi^-$ - spectra}
\label{s10kpic} \indent

   The total experimental background has been  obtained by three
methods. In the first method, the experimental distribution on
effective mass with removed areas of the resonance was approximated
by the polynomial function while this procedure has provided the fit
with $\chi^2$=1 and polynomial coefficient with errors less than
30\%. The second of the angle between $K^0_s$ and $\pi$  for
experimental events randomly mixing method was described in
\cite{mix}. The third type of background has been obtained by
FRITIOF model\cite{fri1}.

The statistical significance of resonance peaks were calculated as
NP /$\sqrt{NB}$, where NB is the number of counts in the background
under the peak and NP is the number of counts in the peak above
background. The analysis of background done by three methods has
shown that there are not observable structure in range of peaks. The
statistical significance of resonance peaks were calculated as
NP/$\sqrt{NB}$, where NB is the number of counts in the background
under the peak and NP is the number of counts in the peak above
background.The values for the mean position of the peak and the
width obtained by using Breit-Wigner fits.

\begin{figure}[ht]
 \epsfysize=150mm
 \epsfxsize=150mm
 \centerline{
 \epsfbox{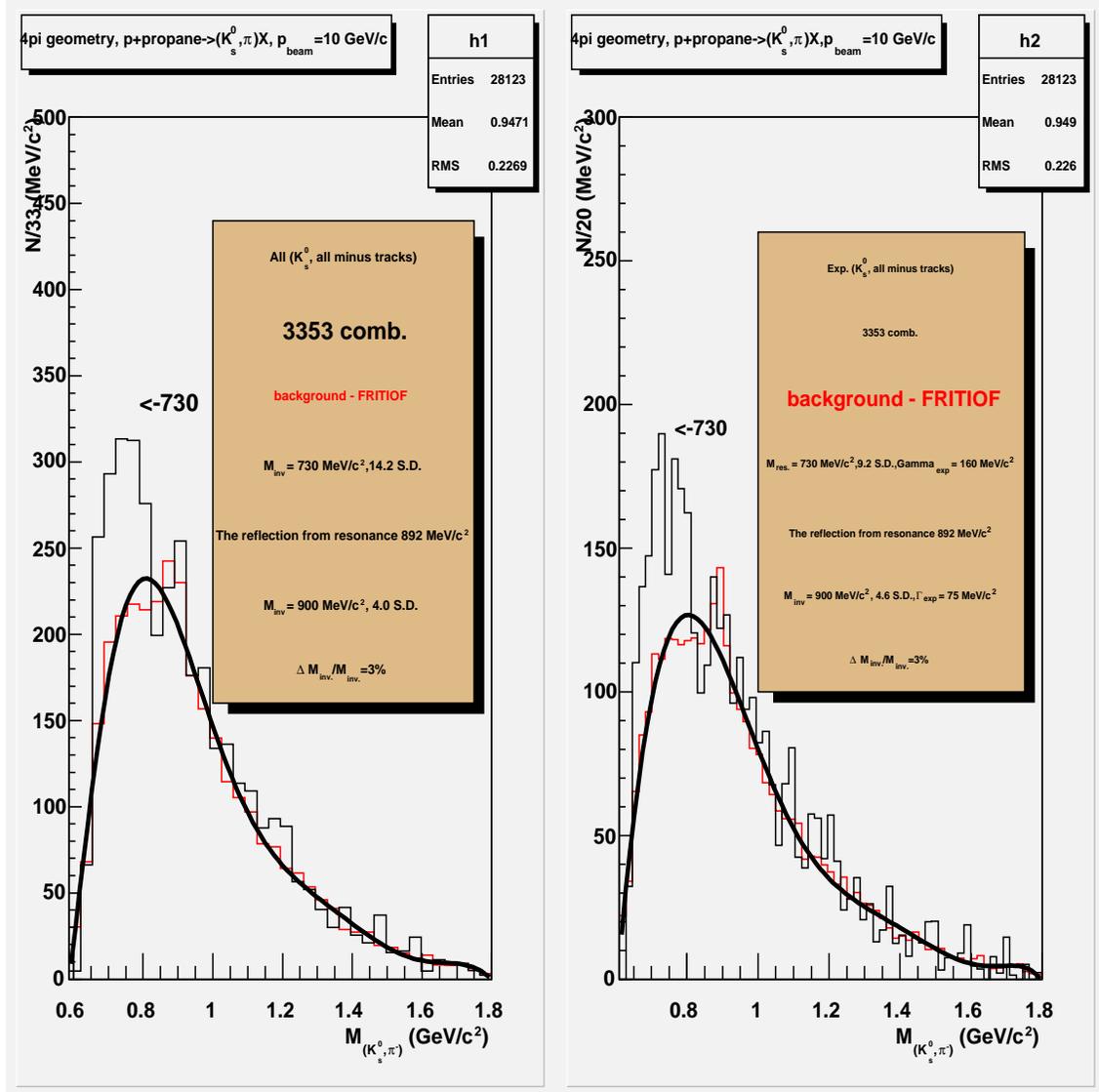}}
 \caption{Invariant mass distribution ( $K^0_s \pi^-$) in the
 inclusive reaction p+$C_3H_8$ . The solid curve is the
 background by the FRITOF taken in the form of a superposition of Legendre
polynomials up to the 7-th degree. The red histogram is the
   background  by FRITIOF  \cite{fri1}. }
  \label{kpi10}
\end{figure}

 Figure~\ref{kpi10}  shows  the invariant mass distribution
 from 3353 ($K^0_s\pi^-$ )combinations with bin sizes 33 MeV/$c^2 $.The
2*$10^6$ events for p+propane $\to\pi^- K^0_s$X interaction by
FRITIOF has simulated.  This background was obtained by using our
experimental condition. In Figure~\ref{kpi10} the simulated
background distribution has been normalized to the experimental
distribution. The solid curve is the  background by the FRITIOF
taken in the form of a superposition of Legendre polynomials up to
the 7-th degree. The values for the mean position of the peak and
the width were obtained by using Breit Wigner fits. There are
significant enhancements in mass regions of 730 and 900 MeV/$c^2$.
The excess above background are equal to 14.2$\sigma$ and 4.0
$\sigma$, respectively. There are small enhancements in mass regions
of (1100-1200) and (1400-1500) MeV/$c^2$. The average effective mass
resolution of $K^0_s\pi^-$ system  is equal to 3.0\%. The peak in
$K^0_s\pi$ invariant mass spectra M(900) is identified as well known
resonances from PDG\cite{pdg} with the mass of 892 MeV/c$^2$.The
background analysis of the experimental data are based on FRITIOF
and the polynomial method  and  defined the behavior of invariant
mass equally.

\begin{figure}[ht]
 \centerline{\epsfysize=150mm,\epsfxsize=150mm
 \epsfbox{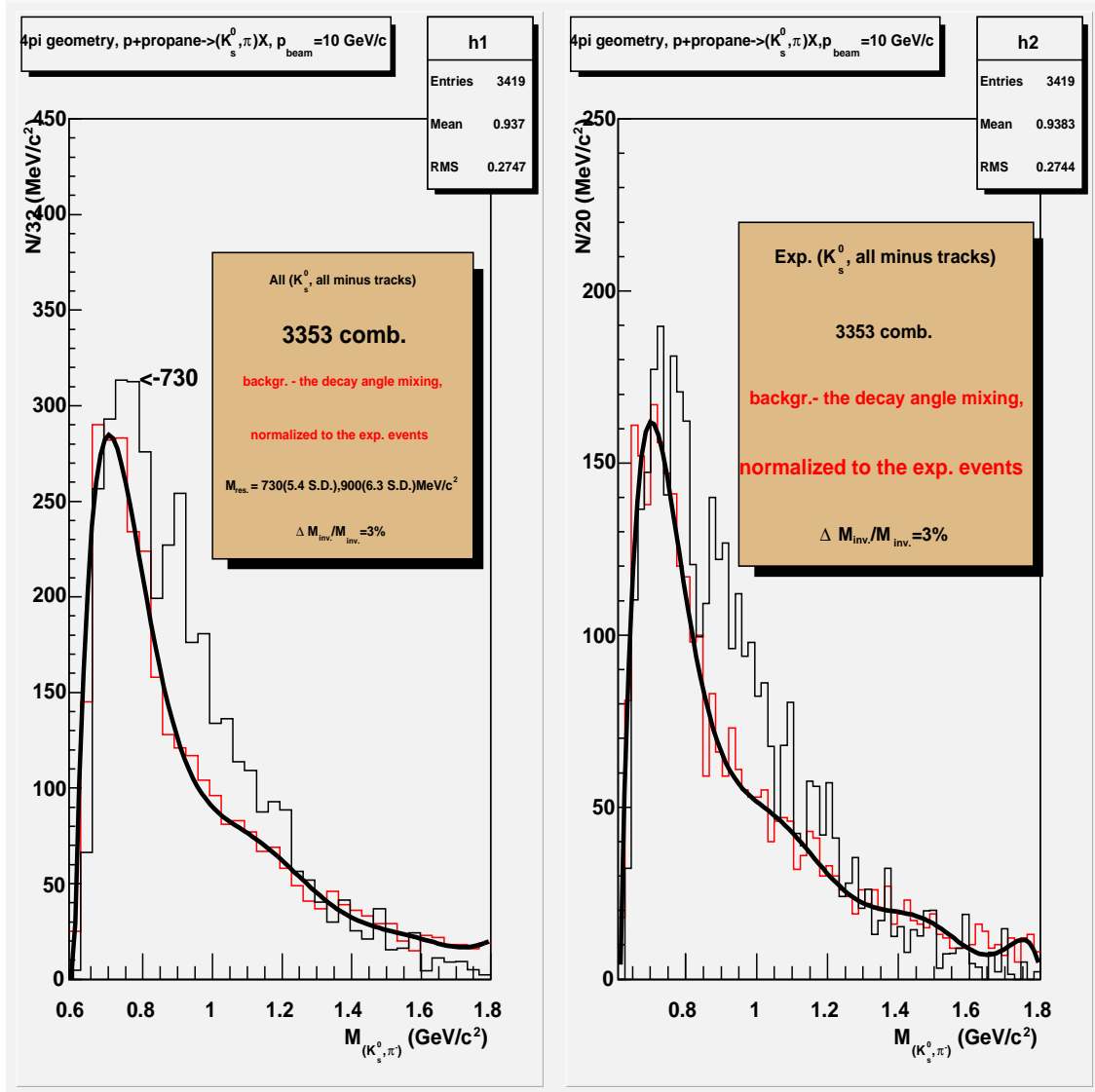}}
 \caption{Invariant mass distribution ( $K^0_s\pi^-$) in the
 inclusive reaction p+$C_3H_8$ . The curve is the  experimental
 background by the secound method taken in the form of a superposition
 of Legendre polynomials up to the 8-th degree.   The red histogram is the
   background  by the mixing method\cite{mix}. } \label{kpi10e}
    \end{figure}

 The effective mass distribution of ($K^0_s\pi$ ) with bin size
  20 MeV/$c^2$ is shown in Figures~\ref{kpi10},\ref{kpi10e}. This bin size
  don't consistent with the experimental resolution. There are statistical
  enhancements in mass regions of 670-750,750-850,850-950, 950-1030 and 1100-1200 MeV/$c^2$

 The background by the mixing methods for p+propane $\to \pi^- K^0_s$X interaction
 has shown in Figure~\ref{kpi10e}.  In Figure~\ref{kpi10e} the
background distribution has been normalized to the experimental
distribution.  Then, the effective mass distribution ($K^0_s\pi^-$ )
was fitted by the 8-th order polynomial
function(Fig.\ref{kpi10e}).There are significant enhancements in
mass regions of 730 and 900 MeV/$c^2$. The excess above background
are equal to 4.6$\sigma$ and 4.2 $\sigma$, respectively.

 \section{Conclusion}

 A number of peculiarities were found in the effective mass spectrum
of system $K^0_spi^-$: 650-850, 850-1050 and 1100-1200 MeV/$c^2$ in
collisions of protons of a 10 GeV/c momentum with propane. The
detailed research of structure of mass spectrum has shown, that the
maximal significant statistical enhancements has been obtained in
730(14.2 $\sigma$) and 900(4.2$\sigma$) effective mass ranges with
$\Gamma \approx$143, 48 MeV/$c^2$ submitted in Table~\ref{res}.

The preliminary total cross section for M(730) production in
p+propane interactions is estimated  to be $\approx 30-250 \mu$b.
 The peak M(730) could be interpreted as a possible candidates for
  a scalar meson state of $\kappa$(800)\cite{kappa}. The M(900) peak
   is interpreted as reflection from well known resonance in
   PDG \cite{pdg} with mass 892 MeV/$c^2$.

\begin{table}
\caption{The effective mass, the width($\Gamma$) and the statistical
significance of resonances produced in collisions of protons with
propane at 10 GeV/c}
 \label{res}
\begin{tabular}{|c|c|c|c|c|c|c|c|}  \hline
Resonance & $M_{\Lambda K^0_s}$&Experimental&&The statistical  \\
Decay & MeV/$c^2$&width $\Gamma_e$&$\Gamma$&significance \\
Mode & &MeV/$c^2$&&$S.D._{max}-S.D._{min}$\\ \hline
  $K^0_s\pi^-$&730&165&143&4.6-14.2\\
 $K^0_s\pi^-$&900&75&48&4.0-6.0\\

 \hline
 \end{tabular}
\end{table}

\newpage
% ----------------------------------------------------------------

\end{document}